\documentclass[pre,twocolumn,aps,eqsecnum]{revtex4}
\usepackage{amsmath,bm,epsfig}

\newcommand{\pa}{\partial}

\begin{document}

\title{Nonequilibrium Thermodynamics of Amorphous Materials I:\\ Internal Degrees of Freedom and  Volume Deformation}

\author{Eran Bouchbinder}
\affiliation{Racah Institute of Physics, Hebrew University of Jerusalem, Jerusalem 91904, Israel}
\author{J. S. Langer}
\affiliation{Dept. of Physics, University of California, Santa
Barbara, CA  93106-9530}

%\date{\today}
\begin{abstract}
This is the first of three papers devoted to the nonequilibrium thermodynamics of amorphous materials. Our focus here is on the role of internal degrees of freedom in determining the dynamics of such systems. For illustrative purposes, we study a solid whose internal degrees of freedom are vacancies that govern irreversible volume changes. Using this model, we compare a thermodynamic theory based on the Clausius-Duhem inequality to a statistical analysis based directly on the law of increase of entropy. The statistical theory is used first to derive the the Clausius-Duhem inequality. We then use the theory to go beyond those results and obtain detailed equations of motion, including a rate factor that is enhanced by deformation-induced noisy fluctuations. The statistical analysis points to the need for understanding how both energy and entropy are shared by the vacancies and their environments.
\end{abstract}

\maketitle

\section{Introduction}
\label{intro}

This is the first of three papers describing our efforts to develop a thermodynamically well-founded theory of nonequilibrium phenomena in amorphous materials. Specific goals of this project are to develop a thermodynamic understanding of the effective disorder temperature and the role that it plays in Shear-Transformation-Zone (STZ) theories of amorphous plasticity \cite{FL98,BLP07I,BLP07II,EB-TEFF-PRE08,JSL-STZ-PRE08}. While working toward these goals, we have encountered a number of fundamental questions. Those questions include: What is the most basic statement of the second law of thermodynamics? How can we reconcile the different approaches to nonequilibrium thermodynamics taken by engineers, applied mathematicians, and physicists? Are the dynamic roles played by internal degrees of freedom properly described by any of those theoretical approaches? Many recent developments in the physics of glassy materials, including the STZ theory, are based on the idea that the state of disorder in such systems is described by an effective temperature that is not necessarily the same as the ordinary temperature. In what sense is the effective temperature a well defined thermodynamic concept?

In this first paper, we focus on questions regarding internal degrees of freedom. For illustrative purposes, we address that issue in the limited context of a simple model of a uniform, not necessarily glassy, solid in which vacancies govern irreversible volume changes. We originally developed this vacancy model as a way of studying irreversible changes in the volume of a glassy material subject to varying temperatures and pressures.  We hope to return to such applications in the future; but, for the present, we use the model purely as an aid for exploring theoretical ideas.  In the second paper \cite{EB-JSL-09-II}, we use the insights gained here to define the effective disorder temperature and to write equations of motion for it. Finally, in the third paper \cite{EB-JSL-09-III}, we reformulate the STZ theory in a way that is consistent with the thermodynamic analysis presented here and in \cite{EB-JSL-09-II}.

Almost by definition, the irreversible responses of materials to applied forces are determined by internal degrees of freedom. Deforming amorphous solids are generally described in terms of internal entities such as flow defects or, of special interest here, STZ's. Qualitatively similar internal structures appear in theories of dislocation motion in crystalline solids, and in nonequilibrium theories of granular materials and complex fluids.  Theories of these dissipative phenomena necessarily invoke the second law of thermodynamics, at the very least as a constraint on the equations of motion for the internal variables.

There is a very large body of literature on this subject. For example, see monographs by Lubliner \cite{LUBLINER-90}, Maugin \cite{MAUGIN-99}, and Nemat-Nasser \cite{NEMAT-NASSER-04}, which we have found to be especially useful. Essentially all of this literature is based on the postulate that the Clausius-Duhem entropy-production inequality is the fundamental statement of the second law of thermodynamics; therefore, we refer to that theoretical starting point as ``conventional''. We recognize, however, that the body of literature to which we are referring contains many different points of view, and that these points of view have continued to evolve in recent decades, especially in the engineering and applied mathematics communities. Our motivation for developing a statistical approach based directly on the law of increase of entropy simply reflects the fact that we have not been able to take any version of the conventional approach far enough to answer the questions that we are asking.

In this paper, we look at the issues concerning the second law and internal degrees of freedom in a simple but physically realistic situation -- the vacancy model mentioned earlier. We start in Sec. \ref{vacancies} by introducing the model and then, in Sec. \ref{conventional}, by briefly summarizing a conventional analysis. The first-principles statistical theory and the resulting nonequilibrium equations of motion are presented in Secs. \ref{statistical} - \ref{EOM}. We conclude in Sec. \ref{conclusions} with some remarks about the broader implications of our results.

\section{Vacancy Model}
\label{vacancies}

We consider a uniform viscoelastic solid of total volume $V$, containing a small but extensive number of vacancies $N_v$. It may be easiest to visualize this solid as being noncrystalline, but that assumption is not essential for present purposes. To avoid the complications of position dependent deformation, we assume that this system remains spatially uniform at all times, and we work with extensive quantities rather than local densities. We further assume that the system is never too far from thermodynamic equilibrium, i.e. that we are not dealing with extremely rapid nonequilibrium phenomena for which local thermodynamic concepts would be invalid. This quasi-equilibrium condition is essential for our arguments in Sec. \ref{statistical}.

Our model, like any model of a material subject only to volume and not shape deformations, is technically viscoelastic rather than elastoplastic. If it is allowed to equilibrate at a fixed temperature and pressure, it eventually returns to the same equilibrium volume with the same number of vacancies; whereas a true elastoplastic material would return to a permanently deformed shape if subjected to shear. A second difference is that a plastic material can undergo steady-state shear flow, and usually exhibits a yield stress that marks the onset of that behavior. No such steady-state deformation can occur in the present case.

The fundamental differences between these kinds of inelastic deformation are important, but are not the central issues to be discussed here. Rather, the model of purely volumetric  deformation is especially useful to us because there is no need for anything other than an additive decomposition of arbitrarily large elastic and inelastic volume changes. No special mathematical efforts are needed to compute the results of complex sequences of deformations; but the separation between elastic and inelastic deformations remains a nontrivial topic of interest as discussed, for example, in \cite{06XBM}.

The volume $V$ in this model consists of three additive components:
\begin{equation}
\label{Vdef}
V = V_0 + V_{el} + V_{in}.
\end{equation}
Here, $V_0$ is a reference volume, determined by the entropy (or temperature). For simplicity, we neglect thermoelastic effects and assume that $V_0$ is just a constant.  The elastic volume, $V_{el}$, is associated with reversible changes in the elastic energy.  An increment $\delta V_{el}$ is a change in the total volume at fixed entropy and fixed $N_v$; that is, it takes place with no change in the internal state of the system.  Our central assumption is that the inelastic volume associated with the vacancies is simply $V_{in} = v_0\,N_v$, where $v_0$ is the effective volume of a vacancy.   $V_{el}$ and $V_{in}$ are independently ``variable'' but not independently ``controllable.''  In equilibrium, $V_{el}$ is controlled directly by the pressure. On the other hand, $N_v$ is a ``hidden'' internal variable so long as the system is not coupled to a chemical-potential reservoir that controls the number of vacancies.  Nevertheless, we must treat $N_v$ as having its own dynamics, and being able at any time to change in ways that are not directly constrained by concurrent changes in $V_{el}$ or the entropy.  Specifically, we assume that the rate of inelastic volume deformation is
\begin{equation}
\label{Nvdot}
\dot V_{in} = v_0\,\dot N_v,
\end{equation}
and that $N_v$ is a dynamical variable that satisfies its own equation of motion.

\section{Conventional Theory}
\label{conventional}

A conventional analysis of this model, described along lines laid out by Coleman, Noll and Gurtin in the 1960's \cite{COLEMAN-NOLL-63,COLEMAN-GURTIN-67}, starts by writing the first law of thermodynamics in the form
\begin{equation}
\label{firstlaw}
\dot U = - p\,\dot V + Q,
\end{equation}
where $U$ is the internal energy, $p$ is the pressure, $-p\,\dot V$ is the work done on the system and $Q$ is the rate at which thermal energy is entering the system. The conventional theory then {\it postulates} that there exists an entropy $S$ and a temperature $\theta$ defined by a continuity  equation. For this spatially uniform system, that equation is simply
\begin{equation}
\label{secondlaw1}
\dot S - {Q\over \theta} = \Sigma.
\end{equation}
Here, the temperature $\theta$ is expressed in energy units ($k_B = 1$); $Q/\theta$ is the rate at which entropy is entering the system; and $\Sigma$ is the entropy source, i.e. the rate at which entropy is being produced. The conventional statement of the second law, the Clausius-Duhem inequality, says that the entropy production rate is non-negative
\begin{equation}
\label{CD}
\Sigma \ge 0.
\end{equation}
These relations are taken to be axiomatic; they do not presuppose any statistical interpretation of $S$ or $\theta$.

Eliminating $Q$ between Eqs. (\ref{firstlaw}) and (\ref{secondlaw1}), we find
\begin{equation}
\label{secondlaw2}
\theta\,\dot S - \dot U - p\,\dot V = \theta\,\Sigma \ge 0,
\end{equation}
which is conveniently rewritten by transforming to the Helmholtz free energy $F(\theta,V_{el},N_v)= U(S,V_{el},N_v) - \theta\,S$, and then performing the partial differentiations
\begin{eqnarray}
\label{secondlaw2a}
\nonumber
\left({\partial F\over \partial \theta}+ S\right)\,\dot\theta  &+&\left({\partial F\over \partial V_{el}}+p\right)\,\dot V_{el} +\left({\partial F\over \partial N_v}+v_0 p\right)\,\dot N_v\cr\\ &=& - \,\theta\,\Sigma\le 0.
\end{eqnarray}
Here we have used Eqs. (\ref{Vdef}) and (\ref{Nvdot}).

In the spirit of Coleman and Noll \cite{COLEMAN-NOLL-63}, we recognize that this expression consists of three separate, independent inequalities because, as discussed above, the time derivatives are unconstrained by each other. We satisfy the first inequality by identifying $S = - \,\partial F/\partial \theta$, thus recovering the familiar thermodynamic relation. The second inequality usually is satisfied by identifying $p = -\, \partial F/\partial V_{el}$ i.e. using the equilibrium relation between $p$ and $V_{el}$. However, if we were interested in thermoviscoelastic effects, then we would satisfy this inequality by writing a dissipative equation of motion for $V_{el}$ of the form
\begin{equation}
\label{dotVel}
\dot V_{el} = - \gamma_{el}\left(\,{\partial \tilde F_{el}\over \partial V_{el}}\right)_{\theta,N_v};~~~\tilde F_{el} = F + p\,V_{el},
\end{equation}
where $\gamma_{el}$ is a non-negative dissipation coefficient. In this way, we would account for the energy dissipation that accompanies the relaxation of the ``viscous pressure'' $p+\pa F/\pa V_{el}$. Since thermoviscoelasticity is not the topic of primary interest here, we simply adopt the equilibrium relation from here on.

Finally, using Eq. (\ref{Nvdot}), we write the third inequality in the form
\begin{equation}
\label{CD1}
-\left(\,{\partial \tilde F_v\over \partial N_v}\right)_{\theta,V_{el}}\,\dot N_v \ge 0, ~~~\tilde F_v = F + p\,v_0\,N_v.
\end{equation}

This is a specific realization of the Clausius-Duhem inequality, Eq. (\ref{CD}). The same result was obtained in Rice's classic 1971 paper \cite{RICE-71} where, however, the free energy $F$ was assumed to be a function of the total deformation rather than the elastic part alone, so that the pressure was missing in the expression for $\tilde F_v$. The inequality of (\ref{CD1}) is satisfied by
\begin{equation}
\label{dotNv}
\dot N_v = - \,\gamma_v\left(\,{\partial \tilde F_v\over \partial N_v}\right)_{\theta,V_{el}},
\end{equation}
where $\gamma_v$ is again a non-negative dissipation coefficient. More generally, any monotonically increasing function of $N_v$ that vanishes where $\partial \tilde F_v/\partial N_v = 0$ can be used on the right-hand side of Eq. (\ref{dotNv}).

Ever since Coleman and Noll introduced their axiomatic version of thermomechanics, physicists (such as JSL) have been impressed by its mathematical elegance, but have worried that it might be incomplete because it does not start with a statistical definition of entropy. It is not clear what statistical interpretation of the entropy $S$ is implied by the preceding equations or, conversely, how the internal energy $U$ might depend on $S$.  For example, it is not obvious in a conventional formulation how to evaluate the free energy $\tilde F_v$ in Eq. (\ref{CD1}).  More importantly, the Coleman-Noll postulates operationally define a temperature as well as an entropy.  In the analysis presented here, we are looking ahead to an effective temperature theory in which there will be two different temperatures -- a situation that seems to be beyond the scope of the conventional axiomatic formulation.

Lastly, we note that the axiomatic approach makes no mention of a thermal reservoir. It seems to us that any theory of this kind ought to include a specific mechanism by which the temperature is controlled. If that mechanism involves coupling to a thermal reservoir, then the theory ought to predict the rate at which heat is flowing between the system and the reservoir. Conversely, the theory should be able to predict what happens if that flow is constrained, as in an adiabatic process. But any coupling to a reservoir disappears when $Q$ is eliminated in Eq. (\ref{secondlaw2}).

\section{Statistical Theory}
\label{statistical}

The basic statistical statement of the second law is that the system as a whole, including any thermal reservoir to which the subsystem of primary interest may be coupled, must move toward states of higher probability, i.e. to states of higher entropy. Although the Coleman-Noll procedure assigns no {\it a priori} statistical significance to the entropy, this principle lies at its heart. In their formulation, however, the principal focus is on spatial heterogeneities. The entropy of the system as a whole increases as heat flows between spatially separated elements, each of which is always in a state of local equilibrium with its own local energy, entropy, and temperature.  It is conceptually easy, albeit mathematically more complicated, to add spatial heterogeneity to the vacancy model. We do not do this explicitly here but, nevertheless, anticipate the need to reinterpret our uniform model as just one element of a larger, spatially inhomogeneous, coarse-grained system.

Our strategy is to start with a statistical definition of entropy, and to introduce a thermal reservoir, but otherwise to stay as close as possible to the conventional analysis. Therefore, in analogy to Eq. (\ref{firstlaw}), we begin by writing the first law in the form
\begin{equation}
\label{firstlaw3}
- p\,\dot V = \dot U + \dot U_R,
\end{equation}
where $U_R = U_R(S_R)$ is the energy of the reservoir as a function of its entropy $S_R$. Similarly, in analogy to Eqs. (\ref{secondlaw1}) and (\ref{CD}),  the second law is
\begin{equation}
\label{secondlaw3}
\dot S_{neq} + \dot S_R \ge 0,
\end{equation}
where $S_{neq}$ is the entropy of a system that is not necessarily in thermal equilibrium.

The main question is what to use for $S_{neq}$.  We propose, with several conditions to be listed below, that the correct choice of this entropy has the form
\begin{equation}
S_{neq}(U,V,\{\Lambda_{\alpha}\}) = \ln\,\Omega(U,V,\{\Lambda_{\alpha}\}),
\end{equation}
where $\Omega(U,V,\{\Lambda_{\alpha}\})$ is a constrained measure of the number of states of the system with energy $U$, volume $V$, and specified values of a set of internal variables $\{\Lambda_{\alpha}\}$.  The $\Lambda_{\alpha}$'s are out of equilibrium if their values are not the ones that maximize $S_{neq}$.  When all of them do maximize $S_{neq}$, i.e. when $\Lambda_{\alpha} = \Lambda_{\alpha}^{eq}$, then we require that the equilibrium entropy $S_{eq}(U,V) = \ln\,\Omega(U,V)$ be accurately approximated by
\begin{equation}
\label{entropy}
{1\over V}\,S_{eq}(U,V) \approx {1\over V}\,S_{neq}(U,V,\{\Lambda_{\alpha}^{eq}\}).
\end{equation}
This approximation must become an equality in the thermodynamic limit, $V \to \infty$. In general, $S_{neq} < S_{eq}$, because the constrained entropy $S_{neq}$ counts fewer states than the unconstrained entropy $S_{eq}$. We require that the difference between these quantities per unit volume become negligibly small as $\{\Lambda_{\alpha}\}\!\to\!\{\Lambda_{\alpha}^{eq}\}$ and as the size of the system becomes indefinitely large. Without this condition, we would not have a single, well defined entropy upon which to base a self-consistent thermomechanical theory.

Validity of Eq. (\ref{entropy}) therefore requires that three conditions be satisfied:

(1) The set of variables $\{\Lambda_{\alpha}\}$ must be sub-extensive.  If there are $N_{\alpha}$ such variables, and there are $N$ total degrees of freedom in the system, then $N_{\alpha}/N$ must vanish in the thermodynamic limit.  Otherwise, the variations of the $\Lambda_{\alpha}$'s would produce an extensive entropic correction to the equilibrium free energy, and Eq. (\ref{entropy}) would not be correct. More explicitly, note that we can compute $S_{eq}$ by integrating over each of the variables $\Lambda_{\alpha}$ in $\Omega(U,V,\{\Lambda_{\alpha}\})$, obtaining a correction to $\ln \,\Omega$ proportional to  $N_{\alpha}$.  That correction must be negligibly small compared to $S_{neq}(U,V,\{\Lambda_{\alpha}^{eq}\})$, which is proportional to $N$.

(2) We must be working in the quasi-equilibrium limit, where all the unconstrained degrees of freedom have rapidly come to equilibrium, and where their fluctuations have been accounted for in computing $S_{neq}$.

(3) Condition (1) requires that the $\Lambda_{\alpha}$'s be coarse-grained variables. If there is only a sub-extensive number of these variables, then each of them must be a sum over a statistically large number of degrees of freedom. That is, the $\Lambda_{\alpha}$ themselves must be extensive. (Of course, nothing prevents us from interpreting them as spatial averages over a macroscopically large system.)  It then follows that the entropies associated with each of the $\Lambda_{\alpha}$ must be included explicitly in $S_{neq}$.  For example, our single internal variable $N_v$ describes an extensive population of vacancies.   The associated entropy, i.e. the logarithm of the number of ways in which the $N_v$ vacancies can be arranged in the volume $V$, must  be contained in $S_{neq}$.

Accordingly, the entropy appearing in Eq. (\ref{secondlaw3}) is
\begin{equation}
S_{neq}(U,V_{el},N_v) = \ln\,\Omega(U,V_{el},N_v).
\end{equation}
For reasons discussed in Sec. \ref{vacancies}, we replace $V$ by $V_{el}$ as an independent argument of $S_{neq}$. We then invert $S_{neq}(U,V_{el},N_v)$ to obtain $U(S_{neq},V_{el},N_v)$.  We identify
\begin{equation}
\label{U-theta}
\left({\partial U\over \partial S_{neq}}\right)_{V_{el},N_v} = \theta
\end{equation}
and, as stated following Eq. (\ref{dotVel}), we use the equilibrium thermodynamic relation for the pressure
\begin{equation}
\label{U-p}
\left({\partial U\over \partial V_{el}}\right)_{S_{neq},N_v} = - \,p.
\end{equation}
Therefore,
\begin{equation}
\dot U = -\,p\,\dot V_{el} + \left({\partial U\over \partial N_v}\right)_{S_{neq},V_{el}}\dot N_v + \theta\,\dot S_{neq}.
\end{equation}
The first law, Eq. (\ref{firstlaw3}), becomes
\begin{equation}
\label{firstlaw4}
-\,p\,\dot V_{in} - \left({\partial U\over \partial N_v}\right)_{S_{neq},V_{el}}\dot N_v - \dot U_R = \theta\,\dot S_{neq},
\end{equation}
where we have used $\dot V = \dot V_{el} + \dot V_{in}$ to eliminate $\dot V_{el}$.

At this point, we depart from the strategy that led to Eq. (\ref{secondlaw2}). Instead of eliminating the coupling to the thermal reservoir as was done there, we use Eq. (\ref{firstlaw4}) to evaluate $\dot S_{neq}$ in the second law, Eq. (\ref{secondlaw3}); and we identify $\dot S_R = \dot U_R/\theta_R$, where $\theta_R = \partial U_R/\partial S_R$ is the reservoir temperature. We also use Eq. (\ref{Nvdot}) to eliminate $\dot V_{in}$ in favor of $\dot N_v$.  The result is
\begin{equation}
\label{secondlaw4}
{\cal W}(p,N_v,\dot N_v) - \left( 1 - {\theta\over \theta_R}\right)\,\dot U_R\ge 0,
\end{equation}
where
\begin{equation}
\label{Wdef}
{\cal W}(p,N_v,\dot N_v)= -\left[\,p\,v_0 + \left({\partial U\over \partial N_v}\right)_{S_{neq},V_{el}}\right]\dot N_v
\end{equation}
is the rate at which inelastic work is done on the system minus the rate at which energy is stored by the vacancies.  As will be seen, ${\cal W}$ is a dissipation rate that appears in various forms throughout this series of papers.

The appearance of $\dot U_R$ in this inequality is important, because we control the temperature of the system by controlling the reservoir temperature.  Thus the inequality in Eq. (\ref{secondlaw4}) must be satisfied for arbitrary variations of $U_R$, independent of whatever else is happening in the system. We also must satisfy this inequality for arbitrary variations of $N_v$. For example, the vacancy population could be relaxing toward an equilibrium value while $U_R$ remains constant.  Therefore, in the spirit of Coleman and Noll, we argue that the only way to satisfy this combined inequality for all possible variations of the system is to enforce two separate, independent inequalities:
\begin{equation}
\label{CD2}
{\cal W}(p,N_v,\dot N_v)\ge 0,
\end{equation}
and
\begin{equation}
- \left( 1 - {\theta\over \theta_R}\right)\,\dot U_R\ge 0.
\end{equation}

The first of these relations is essentially identical to the Clausius-Duhem inequality in Eq. (\ref{CD1}).  The differences are that  we have derived Eq. (\ref{CD2}) from statistical first principles rather than postulated it, and that we know exactly what energy and entropy are involved in it.

The second inequality is satisfied by requiring that $\dot U_R$ be a function of $\theta$ that changes sign only when $\theta = \theta_R$; therefore we write
\begin{equation}
\label{Qdef2}
-\dot U_R = A(\theta,\theta_R)\,(\theta_R-\theta)\equiv  Q,
\end{equation}
where $A(\theta,\theta_R)$ is a non-negative function of its arguments.  Here, $Q$ has the same meaning that it had in  Eq. (\ref{firstlaw}) -- the rate at which heat is flowing into the system, in this case, from the reservoir -- but now $Q$ is a well defined function of $\theta$, and Eq. (\ref{Qdef2}) is an equation, not an inequality. With this definition of $Q$, Eq. (\ref{firstlaw4}) becomes
\begin{equation}
\label{firstlaw5}
 \theta\,\dot S_{neq} = {\cal W}(p,N_v,\dot N_v)+ Q.
\end{equation}

\section{Specifics of the Vacancy Model}
\label{specifics}

Because we have an unambiguous definition of the total entropy, and because we know that the entropy of the vacancies must be included in it, we can write $S_{neq}(U,V_{el},N_v)$ in the form
\begin{eqnarray}
\label{SNv}
\nonumber
&&S_{neq}(U,V_{el},N_v) = S_0(N_v) + S_1(U_1)\cr \\&&= S_0(N_v) + S_1\bigl[U - e_0\,N_v -U_{el}(V_{el})\bigr].
\end{eqnarray}
Equivalently, we can  invert this relation and write it as an expression for the internal energy $U$
\begin{eqnarray}
\label{UNv}
\nonumber
&&U(S_{neq},V_{el},N_v) = U_0(N_v) + U_1(S_1) + U_{el}(V_{el})\cr\\&&= e_0\,N_v + U_1\bigl[S_{neq}-S_0(N_v)\bigr] + U_{el}(V_{el}).
\end{eqnarray}
Here $U_0(N_v)$ is the energy of the vacancies, $e_0$ is the formation energy of a vacancy, $S_0(N_v)$ is the entropy of the  vacancies, $U_{el}(V_{el})$ is the elastic energy, and $S_1$ and $U_1$ are, respectively, the entropy and energy of all the other configurational, kinetic, and vibrational degrees of freedom in the system. The structure of these relations, i.e. the arguments of $U_1$ and $S_1$ in their second versions, describes the way the energy and entropy are shared between the vacancies and the other degrees of freedom. Note that the total entropy and energy in Eqs. (\ref{SNv})-(\ref{UNv}) are assumed to have very simple forms. For example, we have omitted a standard thermoelastic term proportional to $S_1 V_{el}$ in Eq. (\ref{UNv}).

For specificity, we assume that the vacancies are very dilute, so that
\begin{equation}
S_0(N_v) = -\,N_v\,\ln\left({N_v\over N_0}\right) + N_v,
\end{equation}
where $N_0$ is the number of sites at which vacancies might occur. Then, using Eq. (\ref{UNv}), we find that
\begin{eqnarray}
\label{dUdNv}
\left({\partial U\over \partial N_v}\right)_{S_{neq},V_{el}}\!\!\!\!&=&\frac{d}{dN_v}\left[U_0(N_v)\!-\!\theta S_0(N_v)\right]\cr &=& e_0 + \theta\ln\!\left({N_v\over N_0}\right).
\end{eqnarray}

If we write
\begin{equation}
\left({\partial U\over \partial t}\right)_{V_{el},N_v} = \theta\,\dot S_{neq} \equiv C_V\,\dot \theta,
\end{equation}
and interpret the extensive quantity $C_V$ to be the heat capacity at constant volume, then Eq. (\ref{firstlaw5}) becomes
\begin{eqnarray}
\label{Qdef3}
\nonumber
 &&C_V\,\dot \theta + \left[e_0 + p\,v_0+\theta\,\ln\left({N_v\over N_0}\right)\right]\,\dot N_v \cr \\&& = C_V\,\dot \theta + \left({\partial G_v\over \partial N_v}\right)_{\theta,p}\dot N_v = Q,
\end{eqnarray}
where the vacancy-related Gibbs free energy $G_v$ is
\begin{equation}
\label{Gdef}
G_v(\theta,p,N_v) = e_0\, N_v  - \theta \,S_0(N_v)+ p\, v_0 \,N_v.
\end{equation}
The Clausius-Duhem inequality, Eq. (\ref{CD2}), is
\begin{equation}
\label{CD3}
- \left({\partial G_v\over \partial N_v}\right)_{\theta,p}\dot N_v \ge 0.
\end{equation}
Clearly, this term in Eq. (\ref{Qdef3}) is a non-negative rate of heat production associated with the relaxation of the internal variable $N_v$ toward an equilibrium value.

\section{Equation of Motion for $N_v$}
\label{EOM}

At this point in the analysis, the standard procedure is to postulate a general form for an equation of motion for $N_v$, and to use the Clausius-Duhem inequality in Eq. (\ref{CD3}) to constrain the parameters that appear in it.  In the present case, there is no reason why $N_v$ should do anything more complicated than relax toward a stable equilibrium value. Therefore, for small departures from equilibrium, we write
\begin{equation}
\label{dotNv2}
\tau_0\,\dot N_v = \tilde\Gamma(N_v)\,\Bigl[N_v^{eq}(\theta,p) - N_v\Bigr],
\end{equation}
where $\tau_0$ is a time scale, $\tilde\Gamma$ is a positive, dimensionless rate factor that we anticipate will be a function of $N_v$ (as well as $\theta$ and $p$), and $N_v^{eq}(\theta,p)$ is the equilibrium value of $N_v$ at the given temperature and pressure. A convenient alternative form of this equation is
\begin{equation}
\label{dotNv2a}
\tau_0\,{\dot N_v\over N_v} = \tilde\Gamma(N_v)\,\Bigl[{N_v^{eq}\over N_v} - 1\Bigr]\cong -\,\tilde\Gamma(N_v)\,\ln\left({N_v\over N_v^{eq}}\right).
\end{equation}
Having no information about nonlinear corrections to Eq. (\ref{dotNv2}), we can use the second expression on the right-hand side of Eq. (\ref{dotNv2a}) just as well as the first, and will do so from here on. Other nonlinear equations of motion for $N_v$ can easily be incorporated into this analysis when justified by some physical mechanism.

To satisfy the inequality in Eq. (\ref{CD3}), we require that both  $\partial G_v/\partial N_v$, and the expression for $\dot N_v$ on the right-hand side of either Eq. (\ref{dotNv2}) or Eq. (\ref{dotNv2a}), vanish at the same point, i.e. at $N_v =N_v^{eq}(\theta,p)$. Thus $N_v^{eq}(\theta,p)$ is the solution of
\begin{equation}
\left({\partial G_v\over \partial N_v}\right)_{\theta,p,N_v = N_v^{eq}}= e_0 + p\,v_0+\theta\,\ln\left({N_v^{eq}\over N_0}\right) = 0,
\end{equation}
and the equilibrium number of vacancies is proportional to a Boltzmann factor
\begin{equation}
\label{Nveq}
N_v^{eq}(\theta,p)= N_0\,\exp\,\left(-{e_0 + p\,v_0\over \theta}\right).
\end{equation}
The inequality in Eq. (\ref{CD3}) is always satisfied so long as $\dot N_v$ is a monotonically decreasing function of $N_v$, which is required for dynamic stability, and is true for both Eqs. (\ref{dotNv2}) and (\ref{dotNv2a}).

Retaining $Q$ explicitly in this analysis has the added benefit of allowing us to deduce an expression for the rate factor $\tilde\Gamma(N_v)$. The equation of motion for $N_v$, as shown in Eq. (\ref{dotNv2}), is a detailed-balance relation in which the vacancy creation rate is proportional to $N_v^{eq}(\theta,p)$.  Therefore, according to Eq. (\ref{Nveq}), the creation rate automatically contains the appropriate Arrhenius activation factor, and $\tilde\Gamma/\tau_0$ can be interpreted as a dimensionless attempt frequency or, equivalently, a noise strength.

Our experience with the STZ theory of plasticity leads us to write $\tilde\Gamma$ as the sum of two terms:
\begin{equation}
\tilde\Gamma = \rho(\theta) + \Gamma(N_v),
\end{equation}
where $\rho(\theta)$ is the strength of the noise generated solely by thermal fluctuations in the absence of mechanical deformation, and $\Gamma(N_v)$ is the noise strength associated with irreversible deformations, i.e. with nonzero $\dot N_v$.  For present purposes, we could simply set $\rho(\theta) = 1$ and let $\tau_0$ be temperature dependent; but we will need the explicit factor $\rho(\theta)$ for discussing glassy systems in the following papers.

An hypothesis (originally due to Pechenik \cite{PECHENIK}) that has worked well for the STZ theory is that $\Gamma$ is proportional to the total rate per vacancy at which heat is generated by the work done on the system. In the present case, this means that
\begin{equation}
{\theta_0\over \tau_0}\,N_v\,\Gamma(N_v) = \theta\,\dot S_{neq} - Q = {\cal W},
\end{equation}
where $\theta_0$ is an energy, and ${\cal W}$ is the same non-negative dissipation rate that was defined in Eq. (\ref{Wdef}). With this assumption, and with the second form of $\dot N_v$ given in Eq. (\ref{dotNv2a}), Eq. (\ref{Qdef3}) becomes
\begin{equation}
-\,\theta\,N_v\,\ln^2\left({N_v\over N_v^{eq}}\right)\,(\rho + \Gamma )+ \theta_0\,\Gamma\,N_v = 0.
\end{equation}
Solving for $\rho + \Gamma $ (a necessarily non-negative noise strength), we find
\begin{equation}
\tilde\Gamma = \rho+ \Gamma = {\rho(\theta)\over 1 - (\theta/\theta_0)\,\ln^2\left(N_v/N_v^{eq}\right)},
\end{equation}
Thus, the mechanically generated noise enhances the rate factor, possibly quite substantially.  The feature that $\tilde\Gamma$ diverges when $N_v$ is sufficiently far from its equilibrium value simply means that the system is dynamically driven away from such values of $N_v$, and that $\tilde\Gamma$ remains positive at all times.

Putting these pieces of the theory together, we have
\begin{equation}
\label{dotNv3}
{\dot N_v\over N_v} = -\,{\rho(\theta)\over \tau_0}\,{\ln\left(N_v/ N_v^{eq}\right)\over 1 - (\theta/\theta_0)\,\ln^2\left(N_v/N_v^{eq}\right)}.
\end{equation}
Equation (\ref{Qdef3}) becomes
\begin{equation}
\label{Qdef4}
C_V\,\dot \theta + \theta\,\ln\left({N_v\over N_v^{eq}}\right)\,\dot N_v = Q =  A(\theta,\theta_R)\,(\theta_R-\theta).
\end{equation}

The combination of Eqs. (\ref{dotNv3}) and (\ref{Qdef4}) allows us to compute time dependent functions $\theta(t)$ and $N_v(t)$ given any driving force $p(t)$ and reservoir temperature $\theta_R(t)$.  The simplest case is the limit in which the coupling to the reservoir is so strong that $\theta = \theta_R$, and the heat capacity of the reservoir is so large that $\theta_R$ remains a constant independent of how much heat is flowing to or from the system.  This assumption, that the temperature is fixed by coupling to the reservoir, is implicit in most thermodynamic theories, but it is actually a bit subtle.  The quantity $Q$ on the right-hand side of Eq. (\ref{Qdef4}) is undefined in the limit $A \to \infty$, $\theta \to \theta_R$, which means that it can assume whatever value is needed in order to keep $\theta = \theta_R =$ constant.  With $\dot \theta = 0$, Eq. (\ref{Qdef4}) is just the Clausius-Duhem inequality again, now telling us that $Q < 0$, and that -- as in the Kelvin-Planck  statement of the second law \cite{ZEMANSKY-51} -- we are not allowed to convert heat directly into work.

Alternatively, suppose that the process is adiabatic, i.e. $A = 0$.  Then Eq. (\ref{Qdef4}) determines how the temperature rises as the external work is converted into internal heat. In the following paper, we will encounter an intermediate case in which $A$ is small but nonzero.

\section{Concluding Remarks}
\label{conclusions}

Although the statistical analysis described here has been developed primarily for use in the effective-temperature theory of amorphous materials presented in the following paper \cite{EB-JSL-09-III}, the present results already point toward some general conclusions.

We have shown that the Clausius-Duhem entropy-production inequality, when applied to the dynamics of internal degrees of freedom, can be derived directly from a statistical interpretation of the second law of thermodynamics -- but only if the  conditions listed following Eq. (\ref{entropy}) are satisfied.  Perhaps the most important of these conditions is that the internal variables must be a small set of extensive quantities in order that the statistical entropy be well defined for nonequilibrium situations.  This condition, in turn, means that entropies as well as energies associated with the internal variables must be included in any dynamical description of the system.

The quantity appearing in the Clausius-Duhem inequality in  Eq. (\ref{CD3}) is interpreted by Lubliner \cite{LUBLINER-72} as  the ``dissipation associated with the internal variables and their conjugate forces.'' Our analysis suggests a sharper and more physically intuitive interpretation -- that the rate of energy dissipation ${\cal W}$ is the difference between the inelastic power $-p\dot V_{in}\! =\! -p\,v_0\,\dot N_v$ and the rate of change of the {\it free} energy $U_0(N_v)-\theta S_0(N_v)$ that is stored in the internal degrees of freedom.

One example of this difference occurs in Rice's 1971 paper \cite{RICE-71}, where he suggests that his internal variables represent an extensive set of slips on slip planes distributed throughout a polycrystalline material. However, he does not calculate the number of ways in which the total slip can be realized as the sum of many individual slips, and therefore does not include the entropy associated with his internal ``averaging variables'' in dynamical formulas analogous to Eqs. (\ref{Gdef}) and (\ref{CD3}).

These issues persist in the more recent literature and, in our opinion, are quite serious. For example, Anand and Su \cite{ANAND-SU-05} implement something like Rice's picture of frictional slips on multiple slip planes by using a phenomenological, nonlinear, rate-dependent relation between local flow and resolved stresses.  Some memory of past deformation is carried by a ``plastic volumetric strain'' and by a related cohesion parameter that appears in the flow equation; but these are scalar quantities that cannot contain information about the directional history of shear flow.  There is no dynamical yield stress as in STZ theory, nor -- so far as we can tell -- is there any way of using the theory to predict what happens when the loading stresses are removed or reversed, partly because the plastic volumetric strain and the cohesion parameter are scalars, but more importantly because neither are properly constituted internal state variables.  The STZ theory has been developed explicitly to overcome such difficulties.  It is the topic of the third paper in this series \cite{EB-JSL-09-III}.

A related question, which is sometimes raised but not answered in the conventional literature, is what happens when the internal degrees of freedom do not relax in a simple manner. In the present case, our irreversible process can be described by a variational principle; that is, our single internal variable $N_v$ moves downhill in the one-dimensional free energy landscape defined by $G_v(N_v)$ in Eq. (\ref{Gdef}). This picture is generalized in the conventional literature by assuming that a system with multiple internal variables moves downhill in a multidimensional inelastic potential. The resulting fluxes obey what is called a ``normality condition,'' or sometimes a ``generalized normality condition'' \cite{LUBLINER-90} because they are assumed to be perpendicular to surfaces of constant (generalized) potential.

We already know that the picture cannot be so simple for the STZ theory, where increasing shear stress drives the system through an exchange of stability between jammed and flowing steady states. This behavior is discussed in detail in the third paper in this series \cite{EB-JSL-09-III}. More generally, we know that no such energy-minimization principles exist for many open situations, where the system is being persistently driven away from equilibrium, and where there are multiple, coupled, internal state variables.  It seems to us that it will be hard to predict a form for a generalized Clausius-Duhem inequality without starting from a first-principles, fully statistical and dynamical description of such systems.

\begin{acknowledgments}
JSL acknowledges support from U.S. Department of Energy Grant No. DE-FG03-99ER45762.
\end{acknowledgments}


\begin{thebibliography}{99}

\bibitem{FL98} M. L. Falk and J. S. Langer, Phys. Rev. E {\bf 57}, 7192
(1998).

\bibitem{BLP07I} E. Bouchbinder, J. S. Langer and I. Procaccia, Phys. Rev. E {\bf 75}, 036107 (2007).

\bibitem{BLP07II} E. Bouchbinder, J. S. Langer and I. Procaccia, Phys. Rev. E {\bf 75}, 036108 (2007).

\bibitem{EB-TEFF-PRE08} E. Bouchbinder, Phys. Rev. E {\bf 77}, 051505 (2008).

\bibitem{JSL-STZ-PRE08} J. S. Langer, Phys. Rev. E {\bf 77}, 021502 (2008).

\bibitem{EB-JSL-09-II} E. Bouchbinder and J. S. Langer, second paper.

\bibitem{EB-JSL-09-III} E. Bouchbinder and J. S. Langer, third paper.

\bibitem{LUBLINER-90} J. Lubliner, {\it Plasticity Theory}, (Macmillan Publishing Company, New York, 1990).

\bibitem{MAUGIN-99} G. A. Maugin, {\it The Thermomechanics of Nonlinear Irreversible Behaviors}, (World Scientific, Singapore, 1999).

\bibitem{NEMAT-NASSER-04} S. Nemat-Nasser, {\it Plasticity}, (Cambridge University Press, Cambridge, UK, 2004).

\bibitem{06XBM} H. Xiao, O. T. Bruhns and A. Meyers, Acta Mechanica {\bf 182}, 31 (2006).

\bibitem{COLEMAN-NOLL-63} B. D. Coleman and W. Noll, Archive for Rational Mechanics and Analysis {\bf 13}, 167 (1963).

\bibitem{COLEMAN-GURTIN-67} B. D. Coleman and M. E. Gurtin, J. Chem. Phys. {\bf 47}, 597 (1967).

\bibitem{RICE-71} J. R. Rice, J. Mech. Phys. Solids {\bf 19}, 433 (1971).

\bibitem{PECHENIK} J. S. Langer and L. Pechenik, Phys. Rev. E {\bf 68}, 061507 (2003); L. Pechenik , Phys. Rev. E {\bf 72}, 021507 (2005).

\bibitem{ZEMANSKY-51} For example, see the historical disccussion on page 147 of M. Zemansky, {\it Heat and Thermodynamics, Third Edition} (McGraw-Hill, New York, 1951)

\bibitem{LUBLINER-72} J. Lubliner, Int. J. Non-Linear Mech. {\bf 7}, 237 (1972).

\bibitem{ANAND-SU-05} L. Anand and C. Su, J. Mech. Phys. Solids {\bf 53}, 1362 (2005).

%\bibitem{KRONER-60} E. Kroner, Archive for Rational Mechanics and Analysis {\bf 4}, 273 (1960).
%
%\bibitem{LEE-69} E. H. Lee, ASME J. Appl. Mech. {\bf 36}, 1 (1969).

\end{thebibliography}
\end{document}